\documentclass[11pt]{article}
\pdfoutput=1
\usepackage{amsmath, amsfonts, amssymb}
\usepackage{comment}
\usepackage{graphicx}
\usepackage{psfrag}
\usepackage{amsthm}
\usepackage[usenames,dvipsnames,svgnames,table]{xcolor}
\usepackage{enumerate}
\usepackage{tcolorbox}
\usepackage{mathtools}
\usepackage{soul}
\usepackage{subfigure}
\usepackage{mathrsfs}  
\usepackage{a4wide}
\usepackage{booktabs}
\usepackage{tikz}
\usepackage{tikz-cd}
\usepackage{makecell}

\usepackage{color}
\definecolor{dark-gray}{gray}{0.20}
\definecolor{gray}{gray}{0.30}
\definecolor{light-gray}{gray}{0.80}
\definecolor{dark-red}{rgb}{0.7,0,0}
\definecolor{dark-green}{rgb}{0.1,0.4,0}
\definecolor{dark-blue}{rgb}{0.3,0.3,0.7}
\definecolor{light-blue}{rgb}{0.8,0.8,1}
\definecolor{swamp}{RGB}{240, 199, 197}

\usepackage{pifont}
\usepackage{setspace}

\newcommand{\be}{\begin{equation}}
\newcommand{\ee}{\end{equation}}

\def\be{\begin{equation}}
\def\ee{\end{equation}}
\def\bea{\begin{eqnarray}}
\def\eea{\end{eqnarray}}

\newcommand{\dd}{\mathrm{d}}

\def\simleq{\; \raise0.3ex\hbox{$<$\kern-0.75em
		\raise-1.1ex\hbox{$\sim$}}\; }
\def\simgeq{\; \raise0.3ex\hbox{$>$\kern-0.75em
		\raise-1.1ex\hbox{$\sim$}}\; }

\numberwithin{equation}{section}

\usepackage{jheppub} 
\hypersetup{
	colorlinks=true,
	linkcolor=dark-blue,
	citecolor=dark-red,
	urlcolor=dark-green,
	linktoc=page,
	pageanchor=false
}

\title{\centering
Curvature corrections to KPV:\\ Do we need deep throats?
}

\author{Arthur Hebecker$^1$,}
\author{Simon Schreyer$^1$,}
\author{ and Victoria Venken$^1$}

\affiliation{$^1$ Institute for Theoretical Physics, Heidelberg University,\\
	Philosophenweg 19, 69120 Heidelberg, Germany} 
\emailAdd{a.hebecker@thphys.uni-heidelberg.de}
\emailAdd{s.schreyer@thphys.uni-heidelberg.de}
\emailAdd{victoria.venken@gmail.com}

\abstract{
We consider $\alpha'^2$ curvature corrections to the action of an NS5-brane which plays the key role in the metastability analysis of warped anti-D3-brane uplifts by Kachru, Pearson and Verlinde (KPV). Such corrections can dramatically alter the KPV analysis. We find that for the $\alpha'^2$-corrections to be sufficiently small to recover essentially the leading-order KPV potential one needs a surprisingly large $S^3$ radius, corresponding to $g_sM > 20$. In the context of the Large Volume Scenario (LVS) this implies a D3-tadpole of at least $\mathcal{O}(10^3-10^4)$. However, large $\alpha'^2$-corrections do not necessarily spoil the uplift in KPV. Rather, as the curvature corrections lower the tension of the brane, a novel uplifting mechanism suggests itself where the smallness of the uplift is achieved by a tuning of curvature corrections. A key underlying assumption is the existence of a dense discretuum of $g_s$. This new mechanism does not require a deep warped throat, thereby sidestepping the main difficulty in uplifting KKLT and LVS. However, all of the above has to be treated as a preliminary exploration of possibilities since, at the moment, not all relevant corrections at the order $\alpha'^2$ are known.
}

\begin{document}

\makeatletter
\let\old@fpheader\@fpheader

\makeatother

\maketitle


\section{Introduction}\label{intro}

The set-up of Kachru, Pearson and Verlinde (KPV) \cite{Kachru:2002gs} is one of the leading proposals for controlled, spontaneous SUSY breaking in string theory. In KPV, a set of anti-D3-branes at the tip of a Klebanov-Strassler throat polarize into a fluxed NS5-brane, which may then decay to restore SUSY or remain metastable\footnote{The corresponding tunneling transitions are key in defining the lifetime of the non-SUSY vacuum \cite{Frey:2003dm, Freivogel:2008wm}.} at finite radius, ideally in a regime controlled in 10d supergravity. 

Our goal is to study $\alpha'^2$ curvature corrections to the NS5-brane of KPV, which is clearly essential to control metastability.
We will show that such corrections can significantly alter the KPV story. To maintain the validity of the leading-order KPV analysis, a strong lower bound on the radius $\sim \sqrt{g_s M \alpha'}$ of the tip of the throat has to be imposed. Here $g_s$ is the string coupling and $M$ counts the units of 3-form flux on the 3-cycle in the throat. If $g_s M$ is small, violating the above bound, $\alpha'^2$-corrections significantly modify the story of KPV and the possibility of a novel uplifting mechanism suggests itself.

We came to consider $\alpha'^2$ corrections in the KPV set-up by first thinking about corrections to the anti-brane uplift, as an essential ingredient in the KKLT \cite{Kachru:2003aw} and Large Volume Scenario (LVS) \cite{Balasubramanian:2005zx, Conlon:2005ki}. We will start by reviewing this simpler anti-D3-brane setup as a motivation and to provide a simplified version of the more complete NS5-brane story which we develop subsequently.

In recent years, concerns have been raised about the existence of de Sitter vacua in string theory (see e.g.~\cite{Danielsson:2018ztv, Obied:2018sgi}). In particular, a series of recent papers \cite{Junghans:2022exo, Gao:2022fdi, Junghans:2022kxg} has scrutinized the conditions required to achieve a de Sitter vacuum in the LVS using an anti-D3-brane uplift.

An effect which was not considered in \cite{Gao:2022fdi} but was crucial to \cite{Junghans:2022exo, Junghans:2022kxg} derives from the $\alpha'^2$ curvature corrections to the action of the anti-D3-brane responsible for the uplift. They lead to the 
anti-D3-induced potential\footnote{
We believe that $c$ should be $3 \times 1.9747$ rather than $1.9747$ as in \cite{Junghans:2022exo, Junghans:2022kxg}. We obtain $1.9747$ when setting the first derivative of the warp factor $h(\tau)$ to zero after computing the Riemann tensor. However, we obtain three times this result when setting the first derivative of the warp factor to zero only {\it after} the scalar curvature-squared expressions have been computed . The difference arises as the scalar expressions contain terms $h'(\tau)/\tau$, which give a finite contribution at the tip, $\tau \rightarrow 0$. By contrast, these terms are lost when setting $h'(\tau)=0$ too early. A more detailed discussion of curvature computations follows in Sect.~\ref{sec:KPVcorrected} and App.~\ref{sec:curvapp}.
}
\begin{equation}
   V_{\overline{D3}} = \mu_3 \text{e}^{-\phi}\left[ 1 - \frac{(4 \pi^2 \alpha')^2}{384 \pi^2} R_{a \alpha b}^{\quad \alpha} R_{\;\beta}^{a \;\; b \beta}\right] = \mu_3 \text{e}^{-\phi}\left[ 1 - \frac{c}{(g_s M)^2}\right]\,,
   \label{eq:antibranecurvecorr}
\end{equation}
where $a$, $b$ are normal indices and $\alpha$, $\beta$ are tangent indices relative to the brane. References \cite{Junghans:2022exo,Junghans:2022kxg} demanded that the anti-D3-curvature corrections should be small compared to the leading order DBI action. In our opinion, { \eqref{eq:antibranecurvecorr} does not necessarily imply} such a strong constraint, as we now explain:\footnote{ Of course, curvature corrections may affect the compactification through effects independent of the modified D3-tension in \eqref{eq:antibranecurvecorr}. We comment on such effects in Sect.~\ref{shallowuplift}, but much more work is needed if one really wants to gain confidence in a regime where the curvature is partially high.}

Note first that the curvature correction appears in the 4D scalar potential as a term which has the same dependence on the Kahler moduli as the tree-level tension. This curvature correction then does not alter the basic set-up of the LVS uplift. One can absorb the curvature correction into a redefinition of the anti-D3-brane tension, effectively lowering the tension.

As a result, having a curvature correction of the same order of magnitude as the tree-level tension is not dangerous to the LVS. In fact, it is helpful as the entire purpose of the warped throat is anyway to lower the effective tension of the anti-D3-brane.
Reducing the tension through curvature corrections by an $\mathcal{O}(1)$ factor will then slightly reduce the amount of warping needed at the tip of the throat to achieve a de Sitter uplift rather than a runaway.

The curvature correction will only become dangerous when $(g_s M)^2 < c$ since, in this case, the brane gives a negative contribution to the potential and is unable to provide an uplift.

Very intriguingly, if the tension can become negative in this manner, this opens up the prospect for a new uplifting mechanism: We know that for very large $g_s M$ the anti-brane tension is positive and essentially uncorrected by curvature terms. Let us assume that, for small $g_s M$, the anti-brane tension turns negative. Now, given a sufficiently dense discretuum of the parameter $g_s M$, which is likely due to the flux dependence of $g_s$, it should be possible to realize a positive anti-brane tension which is tuned to be hierarchically smaller than its flat-space value. One may then use this mechanism to tune the uplifting term in the potential to achieve a metastable vacuum rather than a runaway. There is no need to rely on a deep warped throat. This mechanism could be used in KKLT or LVS. Crucially, the strong constraints associated with the large D3-tadpole contribution coming from a deep warped throat are avoided.

One may object to this picture: If the $\alpha'^2$-corrections are sufficiently large to make the anti-brane tension negative, surely one should also consider e.g.~$\alpha'^4$-corrections and these might contribute positively to the tension. It is then conceivable that, with all corrections summed up, the anti-brane tension is positive for all $g_s M$. While this may be hard to determine, it is clear that there are two logical possibilities:

First, with all corrections taken into account the anti-D3-brane tension may always remain positive. Corrections shifting the anti-brane tension as in \eqref{eq:antibranecurvecorr} are then never a cause for concern.

Second, the anti-brane tension may become negative for some value of $g_sM$. Corrections like those in \eqref{eq:antibranecurvecorr} are then dangerous and one should demand that they remain sufficiently small. However, it is sufficient to require that the corrections are small enough for the anti-brane tension to remain positive, it is not necessary to demand that they are parametrically small as done in \cite{Junghans:2022exo,Junghans:2022kxg}. Moreover, in this case an alternate uplifting mechanism emerges which does not rely on the deep warped throat but instead tunes the anti-brane potential to be exponentially small because of curvature corrections.

All this being said, the anti-D3-brane does have decay channels. Their existence threatens the optimistic story just developed. The best-understood decay channel arises because $p$ anti-D3-branes puff up into an NS5-brane with $p$ units of 2-form flux, as described by KPV \cite{Kachru:2002gs}. This NS5-brane configuration represents a metastable local minimum for $p/M < 0.08$. If, on the contrary, $p/M > 0.08$, the classical decay corresponding to the annihilation of the anti-D3-branes against the background flux in the throat becomes possible. Clearly, to establish or disprove our optimistic anti-D3-story sketched earlier, it is necessary lift the discussion of $\alpha'^2$ corrections to the technical level of KPV and hence to curvature corrections for the NS5-brane. This is the goal of our paper.

To see more clearly why $\alpha'^2$ corrections are crucial, let us recall some well-known parametric estimates for KPV and the anti-D3-uplift: The warp factor of the throat is $\sim\exp(-8\pi N/3g_sM^2)$, where $N$ is the contribution of the throat to the total D3 tadpole. Hence, $g_sM^2$ is a key phenomenological quantity and it may be constrained as follows: 
First, $p/M < 0.08$ together with $p\geq 1$ imply $M\geq 12$. 
Second, for supergravity control the radius of the $S^3$ at the tip of the throat, which scales like $R_{S^3}\sim \sqrt{g_s M}$, should be large. Implementing this through the condition $g_s M \gg 1$ one finds $g_s M^2 \gtrsim 12$. However, this argument is dubious in that for hierarchical supergravity control one must demand $g_s M \gg 1$, not just $g_s M \gtrsim 1$. If one wishes to use a bound of the form $g_s M \gtrsim \mathcal{O}(1)$, one must compute the actual numerical coefficient for this bound and this is where $\alpha'^2$ corrections become important. Depending on the precise coefficients, the final bound can become either significantly stronger or weaker than $g_s M^2 \gtrsim 12$. Jumping ahead, let us quote some of our results: The conservative requirement of a metastable NS5 configuration in the curvature-corrected analysis suggests $g_s M \gtrsim 20$. It turns out that in this situation the choice $p=2$ is optimal, which together with $g_s<1$ and $p/M < 0.08$ implies $M \geq 25$ and $g_s M^2 \gtrsim 500$. As we shall see, the actual impact of $\alpha'^2$ corrections is more intricate, but the above should suffice to provide a feeling for their importance.

Most optimistically, one may only demand that the barrier preventing the decay of the curvature-corrected NS5 is at positive potential. This leads to $g_sM\gtrsim 4$ and hence, using also $p=1$, $p/M=0.08$ and $M=12$, it implies $g_s M^2 \gtrsim 48$.

While the leading-order KPV analysis suggests $g_s M^2 > 12$, it has been proposed that avoiding a conifold instability for an anti-brane in the Klebanov-Strassler throat requires $g_s M^2 > 46$ \cite{Bena:2018fqc, Blumenhagen:2019qcg, Bena:2019sxm, Randall:2019ent,Scalisi:2020jal}. Recently, the existence of this instability has been questioned \cite{Lust:2022xoq}. However, in light of our results it appears that the curvature-corrected KPV stability bound is much stronger than previously thought, possibly dominating any potential conifold instability issue.

Instead of considering the Klebanov-Strassler throat for the uplift one may consider the S-dual set-up \cite{Gautason:2016cyp} where $K$ units of $H_3$-flux are present on the $S^3$ cycle in the throat. Now the anti-D3-branes puff up into a D5-brane at the tip\footnote{We thank Thomas Van Riet for making us aware of this interesting set-up.}. We expect our analysis to go through analogously in this set-up, with the difference that in Klebanov-Strassler the radius of the tip is set by $\sqrt{g_s M}$ while in the S-dual the radius is set by $\sqrt{K}$. We will consider KPV rather than its S-dual as this is the more commonly discussed set-up, but the $S$-dual has the advantage that we are more certain about curvature corrections.

Finally, it is a key question whether our novel fine-tuned uplifting mechanism will still work, even after lifting the simplified anti-D3 curvature correction analysis to the level of the NS5-brane. A sufficient condition would be that the maximum of the KPV potential is lowered to zero value while remaining in the calculationally controlled regime. Our analysis of expected $\alpha'^2$ terms does indeed show such a behaviour of the potential barrier. But this is clearly insufficient to claim success. From the perspective the curvature-corrected NS5-brane analysis, the novel fine-tuned uplifing mechanism remains only an intriguing possibility. To establish this proposal, much more work is required.

An essential shortcoming of our analysis is that we have only taken into account the pure curvature corrections, which are known for the D5-brane case and which we have adapted to the NS5-brane. We expect further corrections involving flux which may be equally important and which are expected to change crucial numerical coefficients in our results. While we believe that our parametric results in terms of these numerical coefficients are robust, a complete analysis with all terms has to be performed in future work.

\section{Curvature corrections to KPV}
\label{sec:KPVcorrected}

\subsection{Reviewing KPV} \label{sec:KPV}

We start by briefly reviewing the set-up of KPV \cite{Kachru:2002gs}. For this and the computation of the curvature corrections to the NS5-brane, we consider the Klebanov-Strassler throat with metric \cite{Klebanov:2000hb,Herzog:2001xk}
\begin{equation}
    \dd s^2= h^{-1/2}(\tau)\dd x_\mu \dd x^\mu + h^{1/2}(\tau) \dd s_6^2\,,
    \label{eq:KSmetric}
\end{equation}
where $\dd x_\mu \dd x^\mu$ is the metric of 4D Minkowski space and $\dd s_6^2$ the metric of the deformed conifold:
\begin{equation}
    \hspace*{-.2cm}\dd s_6^2= \frac{ K(\tau)}{2} \left[ \frac{\dd\tau^2 +(g_5)^2}{3 K^3(\tau)} +\cosh^2(\tau/2)((g_3)^2 + (g_4)^2)+\sinh^2(\tau/2)((g_1)^2 + (g_2)^2)\right].\hspace{-.2cm}
    \label{eq:conifold}
\end{equation}
Here, the one-forms $g^{1..5}$ (see e.g.~\cite{Herzog:2001xk}) parametrize the five directions of the $T^{1,1}$, which can be viewed as an $S^2$ fibration over $S^3$, and
\begin{equation}
    K(\tau) = \frac{(\sinh(2\tau)-2\tau)^{1/3}}{2^{1/3}\sinh(\tau)}\,.
\end{equation}
The warp factor is given by
\begin{equation}
    h(\tau) = \left(g_s M \alpha' \right) 2^{2/3} I(\tau)\,, \qquad I(\tau) =
    \int_\tau^\infty \dd x \,\frac{x \coth(x)-1}{\sinh^2(x)}\, \left(\sinh(2x)-2x\right)^{1/3}\,.
    \label{eq:warpfactor}
\end{equation}
Here $M$ counts the units of $F_3$-flux on the A-cycle of the KS throat.

If $p$ anti-D3-branes are placed at the tip of this throat, they can puff up into an NS5-brane with $p$ units of worldvolume flux. This NS5-brane wraps an $S^2$ inside the $S^3$ whose metric can be written as
\begin{equation}
    R_{S^3}^2\, \dd\Omega_3^2 = R_{S^3}^2\left( \dd\psi^2 + \sin^2(\psi) \, \dd\Omega_2^2\,\right)\,,
\end{equation}
where\footnote{We work in units where $\alpha'=1$ except where we explicitly display $\alpha'$ for clarity.} $R_{S^3}= b_0 \sqrt{g_s M }$ and $b_0^2= 2 a_0^{1/2} / 6^{1/3}\approx 0.93266$ with $a_0=I(0)\approx 0.71805$.

As realized by KPV \cite{Kachru:2002gs}, this opens up a potential decay channel: The NS5 can slip over the equator, with its anti-D3 charge annihilating against 3-form flux. Eventually, the NS5 turns into $M-p$ D3-branes and supersymmetry is restored. The key quantity determining whether this decay process can take place is $p/M$: For small enough values of $p/M$, a meta-stable minimum below the equator exists.

\begin{figure}
    \centering
    \includegraphics[width=1\textwidth]{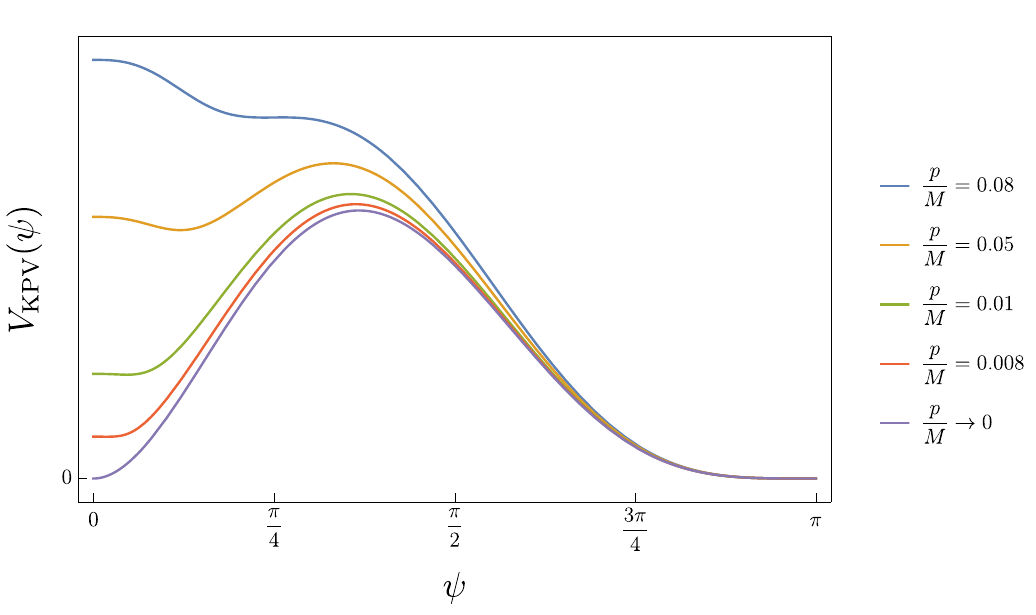}
    \caption{The potential $V_{\text{KPV}}(\psi)$ (suitably normalized) for different values of $p/M$.}
    \label{fig:KPVpot}
\end{figure}

KPV quantify this by deriving the potential $V(\psi)$ of the NS5 at leading order in $\alpha'$ (cf.~Fig.~\ref{fig:KPVpot}):
\begin{equation}
    \hspace*{-.2cm}V_{\text{KPV}}(\psi) = \frac{4 \pi^2 p \mu_5}{g_s} + \frac{4 \pi \mu_5 M}{g_s}\left(\sqrt{b_0^4 \sin^4 (\psi) + \left(p \frac{\pi}{M} -\psi +\frac{1}{2}\sin(2\psi) \right)^2} -\psi +\frac{1}{2}\sin(2\psi) \right)\hspace*{-.2cm}\,.
    \label{eq:kpv}
\end{equation}
In the small $\psi$ expansion of \eqref{eq:kpv}, the meta-stable minimum sits at $\psi_\text{min} \approx 2\pi p/(b_0^4M)$. It disappears when $p/M\gtrsim 0.08 \equiv (p/M)_\ast$. At this value, the minimum and the maximum merge to form an inflection point.

The radius of the NS5 at the minimum is given by $R_{\text{NS}5}\approx 2\pi(p/M) \sqrt{g_sM}/b_0^3$. By increasing $g_sM$ at fixed value of $p/M$, this radius can be made large in string units. Note, however, that this is not the same as making $M$ large while keeping $p$ fixed: In this limit, $R_{\text{NS}5}$ approaches the string length such that the solution can not be trusted.

Let us make two comments concerning this regime of small $p/M$ at large $S^3$ radius $\sim \sqrt{g_sM}$: First, this regime is of great interest since, if one uses the KS throat to uplift an AdS vacuum of type IIB supergravity, the $\overline{\rm D3}$ tension is warped down by $\exp(-8\pi N/3 g_sM^2)$. Here $N$ is the total D3 charge of the throat. Hence, to have large warping without excessive $N$, one wants $M$ not to become too large. As a result, one wants $p$ to be as small as possible.

Second, this regime is under partial control even though the NS5 radius at the minimum is small. Namely, the maximum of $V_\text{KPV}(\psi)$ still corresponds to large NS5 radii and is hence trustworthy.
In fact, as can be seen in Fig.~\ref{fig:KPVpot}, independently of the value of $p/M$, the position of the maximum is quite stable and only increases slightly as $p/M$ decreases. The minimal value of the maximum $\psi_\text{max}^\ast$ (corresponding also to the inflection point) can be approximated by $\psi_\text{max}^\ast=\left.\psi_\text{min}\right|_{(p/M)_\ast} \approx 0.58$. For $p/M\to\infty$ the position of the maximum approaches $\psi_\text{max}\approx1.154$. Thus, we can be sure of the maximum and hence of the existence of a minimum to the left of it, even if the minimum itself can not be controlled.

{ Since our discussion will remain at the probe level, we briefly comment on the regime of validity of the probe approximation: At the very least, the radius $R_{\text{back}}$ of the gravitational backreaction of any brane we consider should be small compared to the $S^3$ radius $\sim \sqrt{g_sM\alpha'}$. The condition for the metric deformation near a $k$-dimensional brane (a `$(k\!-\!1)$-brane') in $d$ dimensions to be small reads $T_k/M_{10}^{d-2} R^{d-k-2} \ll \mathcal{O}(1) $, where $T_k$ is the tension, $1/M_{10}^{d-2}$ the coupling constant, and $R$ the distance to the brane. Solving this for $R$ in the case of $p$ anti-D3-branes gives $R_{\text{back}}^2\sim \sqrt{pg_s}\alpha'$ and hence the condition for backreaction to be small reads
\begin{equation}
    \sqrt{pg_s} \ll g_sM \qquad\qquad \mbox{or} \qquad\qquad \frac{p}{M} \ll g_s M\,.
    \label{eq:probeapprox}
\end{equation}
This can be compatible with large $p/M$. Moreover, one can perform the same analysis for the backreaction of an NS5-brane with $p$ units of worldvolume gauge flux. One reads off from \eqref{eq:kpv} that the tension of such a fluxed NS5 at $\psi \sim {\cal O}(1)$ and $p\gg M$ is $\sim (p/g_s^2 M)\alpha'^{-3}$. This leads to the same bound \eqref{eq:probeapprox} on the ratio $p/M$.
}

\subsection{Curvature corrections} \label{sec:curvature}

The KPV analysis of Sect.~\ref{sec:KPV} can be substantially affected by $\alpha'$ corrections if the curvature at the tip of the throat is not very small. As discussed below, such corrections are suppressed by $1/R_{S^3}^4\sim1/(g_s M)^2$, which for phenomenological reasons can not be taken to zero with arbitrary precision.

Let us start by recalling the curvature corrections to the string-frame DBI action of a D$p$-brane \cite{Bachas:1999um, Junghans:2014zla}. For a constant dilaton, the D$p$-brane DBI action with the $\alpha'^2$ curvature corrections included at leading order in $g_s$ reads
\begin{equation}
\begin{split}
    S_{\text{curv, D}p} =  -\frac{\mu_p}{g_s} \int_{\mathcal{M}_{p+1}} &\dd^{p+1}\xi \, \sqrt{-\det\left(g_{\mu\nu}+ \mathcal{F}_{\mu\nu}\right)}
    \biggl(1- \frac{(2\pi)^4 \alpha'^{\,2}}{24\cdot 32 \pi^2}\biggl[  (R_T)_{\alpha\beta\gamma\delta}(R_T)^{\alpha\beta\gamma\delta} \\&-2 (R_T)_{\alpha\beta}(R_T)^{\alpha\beta} - (R_N)_{\gamma\delta a b}(R_N)^{\gamma\delta a b} + 2 \overline{R}_{ab}\overline{R}^{ab}\biggr]\biggr)\,,
   \end{split} 
    \label{r2}
\end{equation}
where $\mu_p$ is the brane tension, and $\mathcal{F}_{\mu\nu}= B_{\mu\nu} + 2\pi\alpha' F_{\mu\nu}$, with $F$ the worldvolume field strength. The detailed definitions of the various curvature-squared terms and their indices are given in App.~\ref{sec:curvapp}. 

We could now consider S-dualized Klebanov-Strassler (SDKS) \cite{Gautason:2016cyp}, a warped throat geometry in IIB string theory where one has $K$ units of $H_3$-flux on the $S^3$ which remains of finite size at the tip of the throat. The $p$ anti-D3-branes can then puff up into a D5-brane wrapping an $S^2$ on the $S^3$ at the tip of the throat, entirely analogously to how in KPV the anti-D3-branes puff up into an NS5. We shall instead discuss our results in terms of the KPV set-up in the KS throat.
As we will see, we expect the leading curvature corrections to be of the same form and at the same order in $g_s$ compared to the leading term in the DBI action for both the D5-brane and the NS5-brane. We then expect our analysis to be analogous for either the NS5-brane in KPV or the D5-brane in the S-dual set-up, with the difference that in the Klebanov-Strassler throat the tip radius is set by $\sqrt{g_s M}$ while in the S-dual it is set by $\sqrt{K}$.

What are the curvature corrections on the NS5-brane? One may obtain curvature corrections by S-dualizing the D5-brane action from \eqref{r2}, i.e.~replacing $g_s\to 1/g_s$. This yields, after Weyl rescaling $g_{\mu\nu}\to g_{\mu\nu}/g_s$,
\begin{equation}
\begin{split}
  S_{\text{curv, NS}5 \text{ S-duality}} = & -\frac{\mu_{5}}{g_s^2} \int_{\mathcal{M}_{6}} \dd^{6}\xi \,  \sqrt{-\det\left(g_{\mu\nu}+  2\pi g_s F_{2\, \mu \nu} - g_s C_{2 \, \mu \nu}\right)}\, \biggl(1- g_s^2 \frac{(2\pi)^4 \alpha'^{\,2}}{24\cdot 32 \pi^2}\biggl[\\&  (R_T)_{\alpha\beta\gamma\delta}(R_T)^{\alpha\beta\gamma\delta} -2 (R_T)_{\alpha\beta}(R_T)^{\alpha\beta} - (R_N)_{\gamma\delta a b}(R_N)^{\gamma\delta a b} + 2 \overline{R}_{ab}\overline{R}^{ab}\biggr]\biggr)\,.
    \end{split}
    \label{rNS5sdual}
\end{equation} 
We see that these curvature corrections are $g_s^2$ suppressed compared to the leading-order DBI action, while the D-brane curvature terms were leading order in $g_s$. One expects such $g_s^2$ corrections for gravitons from loop corrections in the worldvolume field theory on the NS5-brane \cite{Gao:2022uop}. By contrast, the curvature corrections on the D5-brane arise at the disk level and are enhanced by a factor $1/g_s$ compared to 1-loop expectations. This leads to a conundrum, as we explain in the two following paragraphs:

Consider a fivebrane (either D5 or NS5) with geometry $\mathbb{R}^{1,3}\times S^2$ and $p$ units of worldvolume flux on the 2-sphere. 
One can consider this brane either in flat 10D space or in the KS throat with the NS5-brane wrapping an $S^2$ in the A-cycle or the D5-brane wrapping an $S^2$ in the B-cycle. The analysis is the same in both scenarios as the curvature of the background becomes negligible relative to that of the $S^2$ once the latter is taken to be small. Now consider shrinking the $S^2$ to zero size. The resulting object is a threebrane with the same quantum numbers as $p$ (anti)-D3-branes. We then assume in line with KPV that this object must be a stack of $p$ anti-D3-branes.

Next, consider our fivebrane action as we shrink the $S^2$ radius $R_{S^2}$ to zero size. We look first at the DBI action at leading order in $\alpha'$. As $R_{S^2}\rightarrow 0$, one has $\int_{S^2}\sqrt{g- B_2}\rightarrow 0$ and $\int_{S^2}\sqrt{g- g_s C_2}\rightarrow 0$, while one always has $(2\pi/g_s)\int_{S^2} F_2 = 4 \pi^2 p/g_s$ for the D5 and $(2\pi/g_s^2)\int_{S^2} g_s F_2 = 4 \pi^2 p/g_s$ for the NS5-brane. We then see that for both fivebranes this limit reproduces the leading order action of  anti-D3-branes, \eqref{r2}. Now consider the $\alpha'^2$ terms on the fivebranes as one sends $R_{S^2}\rightarrow 0$. Clearly for the D5-brane, \eqref{r2}, one reproduces the $g_s$ scaling of the $R^2$ terms for the anti-D3-branes. That the precise index structure for the $R^2$ terms of the D5 is such that it matches onto the result for a stack of $p$ anti-D3-branes is something we assume for consistency of the theory. We will comment more on this in Sect.~\ref{sec:outlook}. However, when we perform the $R_{S^2}\rightarrow 0$ analysis for the NS5-brane, \eqref{rNS5sdual}, we see that the resulting $\alpha'^2$ action is $g_s^2$ suppressed compared to the result required for the anti-D3-branes. Given that by assumption we must reproduce $p$ anti-D3-branes from the NS5-brane when sending $R_{S^2}\rightarrow 0$, we conclude that the NS5-brane action must also have $R^2$ terms at leading order in $g_s$. Which index structure should this leading term have? Since we asserted that for the D5-brane this index structure should be given by the $\alpha'^2$ term of \eqref{r2} and the limit $R_{S^2}\rightarrow 0$ operates entirely analogously for the NS5 and D5-brane, we propose that in order to obtain the correct anti-D3-brane limit the leading order in $g_s$ curvature correction for the NS5-brane should have the same index structure as that for the D5-brane. We then have
\begin{equation}
\begin{split}
  S_{\text{curv, NS}5} = & -\frac{\mu_{5}}{g_s^2} \int_{\mathcal{M}_{6}} \dd^{6}\xi \,  \sqrt{-\det\left(g_{\mu\nu}+  2\pi g_s F_{2\, \mu \nu} - g_s C_{2 \, \mu \nu}\right)}\, \biggl(1- \frac{(2\pi)^4 \alpha'^{\,2}}{24\cdot 32 \pi^2}(1+g_s^2)\biggl[\\&  (R_T)_{\alpha\beta\gamma\delta}(R_T)^{\alpha\beta\gamma\delta} -2 (R_T)_{\alpha\beta}(R_T)^{\alpha\beta} - (R_N)_{\gamma\delta a b}(R_N)^{\gamma\delta a b} + 2 \overline{R}_{ab}\overline{R}^{ab}\biggr] +\mathcal{O}(g_s)\biggr)\,.
    \end{split}
    \label{rNS5}
\end{equation} 
In what follows we shall only consider this NS5-brane action at leading order in $g_s$. Note that after S-duality this leading order term should contribute to the (unknown) $g_s^2$ suppressed, 2-loop term for the D5-brane. Such a term is indeed expected to arise for gravitons from field theory loops in an effective field theory on the D5-brane worldvolume \cite{Gao:2022uop}, analogously to how we also expected such $g_s^2$ suppressed corrections to arise on the NS5-brane worldvolume. This provides a further argument that such a leading order in $g_s$ curvature term on the NS5-brane should exist. We finally note that a 1-loop $R^2$ term is also expected to be present on the D5. It would dualize to an ${\cal O}(g_s)$ $R^2$ on the NS5, as displayed at the very end of \eqref{rNS5} above.

\subsection{Curvature-corrected KPV potential and conservative bounds on uplifts}
\label{sec:curvcorrNS5pot}
To find the contribution of the NS5-brane to the 4d potential, we evaluate the curvature correction in \eqref{rNS5} as explained in more detail in App.~\ref{sec:curvapp} and use the values $I(0)=a_0 $, $I'(0)=0$, and $I''(0)=- 2^{2/3}/3^{4/3}$.  
Keeping only the leading order in $g_s$, one then has to integrate these terms over the $S^2$ wrapped by the NS5 which yields (this integral is also calculated in \cite{Kachru:2002gs})
\begin{equation}
\begin{split}
    V_{\text{curv}}(\psi) =&-\frac{1}{(g_s M)^2}\left( c_1 + 2c_2 \cot^2\psi+c_2\cot^4\psi\right)\, \frac{\mu_5}{g_s^2}  \int_{S^2} \sqrt{g+2\pi g_s \mathcal{F} }\\
    =& -\frac{1}{(g_s M)^2}\left( c_1 + 2c_2 \cot^2\psi+c_2\cot^4\psi\right)\frac{4 \pi \mu_5 M}{g_s} \\
     &\quad\quad\times \sqrt{b_0^4 \sin^4 (\psi) + \left(p \frac{\pi}{M} -\psi +\frac{1}{2}\sin(2\psi) \right)^2}\,,
    \end{split}
    \label{eq:vcurv}
\end{equation}
where $c_1\approx8.825$ and $c_2\approx1.891$.
The analysis of KPV should now be redone including \eqref{eq:vcurv}, i.e.~by considering
\begin{equation}
    V_{\text{tot}}(\psi) = V_{\text{KPV}}(\psi) + V_{\text{curv}}(\psi)\,.
    \label{eq:vtot}
\end{equation}
The task is to determine how the KPV bound $g_s M^2 > 12$ (which follows from $p/M \geq 0.08$ together with $g_s M >1$) is modified. 

\begin{figure}[h]
    \centering
    \includegraphics[width=\textwidth]{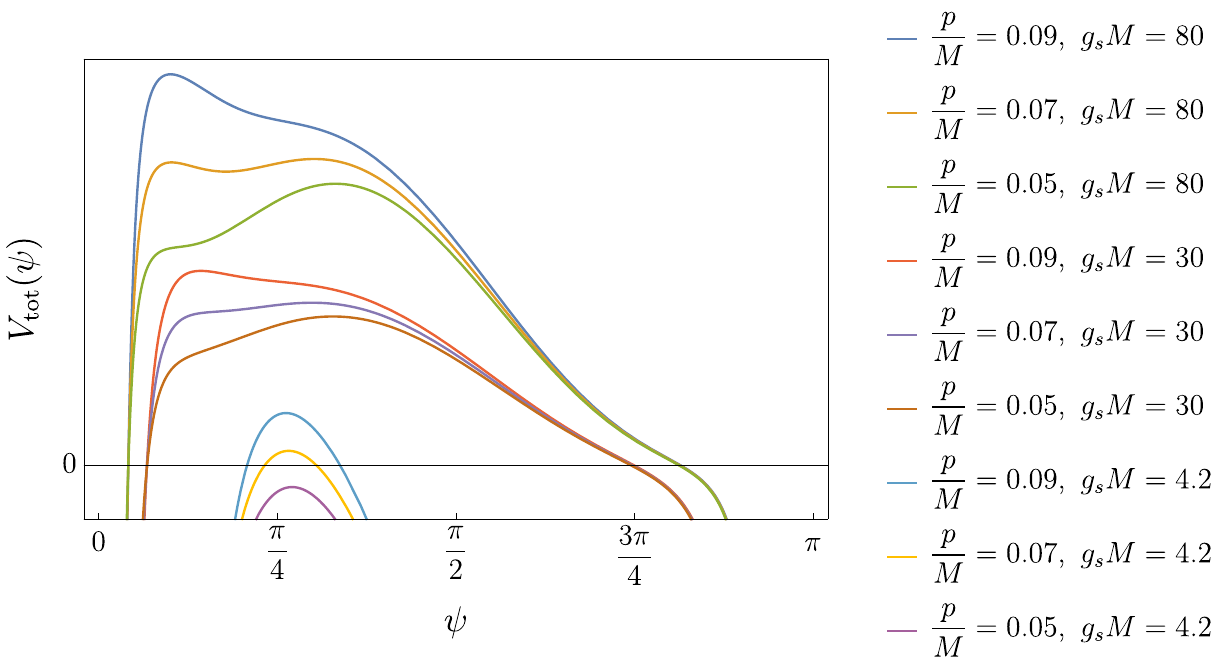}
    \caption{The potential $V_\text{tot}=V_{\text{KPV}}(\psi)+V_{\text{curv}}(\psi)$ for different values of $p/M$ and $g_sM$ in units $\mu_5=1$. For each set of values of $g_sM$, $g_s$ is chosen such that the potential is suitably normalized.}
    \label{fig:KPVpsiplot}
\end{figure}

In the analysis of KPV, the only parameter determining whether a metastable minimum exists is the combination $p/M$. By including the correction \eqref{eq:vcurv}, there are now two parameters which determine the shape of the potential $V_\text{tot}(\psi)$: $p/M$ and $g_s M$. In addition, $V_\text{tot}(\psi)$ contains an overall factor $1/g_s^2$ which sets the overall magnitude of the potential but does not affect its shape\footnote{Interestingly, the corrections to KPV due to conifold dynamics also only depends only on $p/M$ and $g_s M$ \cite{Scalisi:2020jal}, suggesting that there may be a convenient way to deal with corrections to KPV due to conifold dynamics and due to curvature simultaneously. We thank Vincent Van Hemelryck for making us aware of this point.}. 

The potential $V_\text{tot}(\psi)$ is plotted for some sample values of $p/M$ and $g_s M$ in Fig.~\ref{fig:KPVpsiplot}. 
Besides the fact that the original and corrected KPV potential may or may not have a metastable minimum, the corrected potential has a very different appearance when compared to the original KPV result in Fig.~\ref{fig:KPVpot}.
The most notable difference to KPV is that $V_\text{tot}(\psi)$ always has a maximum and always diverges to $-\infty$ as $\psi \rightarrow 0,\pi$. 
These features are due to the extrinsic curvature corrections, which are dominant when the $S^2$ wrapped by the NS5-brane shrinks to a small sphere as they introduce terms $\sim \cot^2\psi$ and $\sim \cot^4\psi$ in the potential. Not to overrate these features, it is crucial to be aware of the limitations of our analysis.

We have only considered $\alpha'^2$ corrections. As the NS5-brane shrinks, corrections at higher orders in $\alpha'$ become increasingly important. Because of this, we can only trust our analysis in the central region of the potential. Further, even at order $\alpha'^2$ we have only accounted for those pure curvature corrections which we know, ignoring completely the effect of flux at order $\alpha'^2$. We expect that these additional corrections prevent any divergent behaviour in the potential. We also assume that, after taking all corrections at all orders in $\alpha'$ into account the potential at $\psi=0$ matches the (all-orders corrected) potential for $p$ nonabelian anti-D3-branes. Similarly, the potential at $\psi=\pi$ should match the (all-orders corrected) potential for $M-p$ nonabelian D3-branes.\footnote{When saying `all orders' we mean including non-perturbative effects.}

At intermediate values of $\psi$, the all-orders-corrected potential may or may not feature a local maximum and metastable minimum depending on the parameters. Note that unlike for $V_\text{KPV}$, the potential at $\psi=\pi$ need not be zero. This is due to the $\alpha'$ terms in the action of the $M-p$ D3-branes left over in the throat after the $p$ anti-branes have annihilated against flux.

\begin{figure}[h]
    \centering
    \subfigure[]{\includegraphics[width=0.47\textwidth]{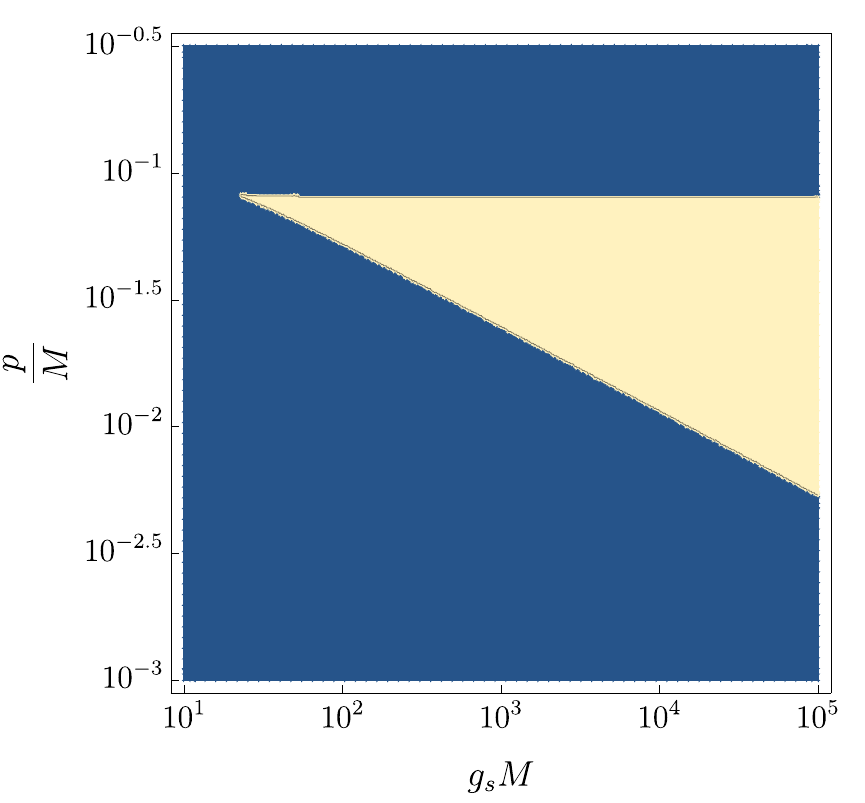}} 
    \subfigure[]{\includegraphics[width=0.46\textwidth]{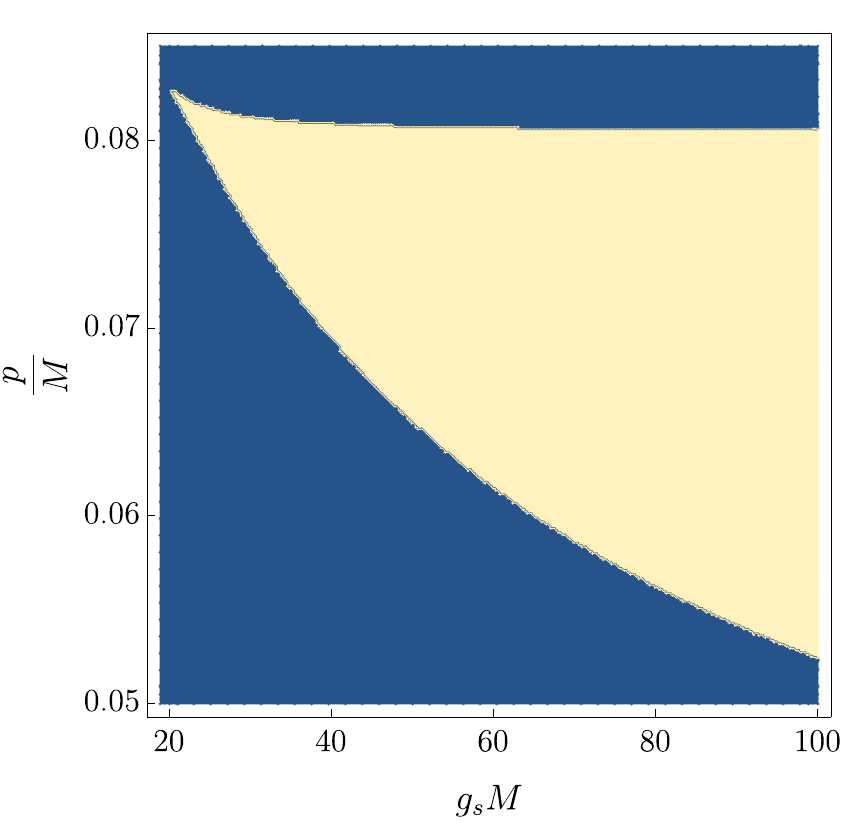}} 
    \caption{The yellow area shows the region in the $(g_sM,p/M)$-parameter space where a meta-stable minimum exists.	(a) is a log-log-plot for a wide range of parameters and (b) zooms into the region of smallest allowed values of $g_sM$.}
    \label{fig:poverm}
\end{figure}

In Fig.~\ref{fig:poverm} we show for which ranges of $p/M$ and $g_s M$ the potential $V_\text{tot}(\psi)$ has a metastable minimum. The tip of this shark fin region is at $p/M=0.0826$ and $g_sM=20.363$. The upper horizontal bound is a slightly weakened version of the KPV bound $p/M <0.08$. This weakening is expected since $V_\text{curv}$ always possesses a maximum that facilitates the existence of the local KPV minimum at smaller values of $\psi$. The horizontal line asymptotes to agree with KPV precisely as $g_sM$ grows. 

The second, diagonal line bounding the shark fin from below corresponds to the $\alpha'^2$ curvature corrections being significant enough to destroy the metastable minimum through a runway to $\psi=0$.\footnote{In the interval $g_sM\in[20.363,100]$, this part of the bound can be approximated by
\begin{equation}
\label{eq:poverm}
    \frac{p}{M}(g_sM) = 0.0313227  + \frac{2.73449}{g_sM} - \frac{72.7164}{(g_sM)^2} + \frac{1182.92}{(g_sM)^3}- \frac{8223.59}{(g_sM)^4} +\mathcal{O}\left((g_sM)^{-5}\right)\,.
\end{equation}}
Crucially, the loss of a metastable minimum at order $\alpha'^2$ does in this case not imply that the fully corrected potential has no metastable minimum. It is just that this minimum may be outside the controlled regime. This may also be the reason why the bound on $g_sM$ is very large, $g_sM>20$. For the metastable minimum at $\psi_\text{min}\sim \sqrt{g_sM}p/M$ to appear in the controlled regime, it needs to be pushed to large values of $\psi$ which requires large $g_sM$. In view of the uplift to dS this is problematic, which can be seen as follows: Combined with the perturbativity condition $g_s<1$, the bound $g_sM>20$ implies $M>20$. For $p=1$, i.e.~$p/M>0.05$, this forces us to even higher $g_sM$, as can be clearly seen from Fig.~\ref{fig:poverm}. In fact, to stay within the shark fin region, one has to use (at least) $p=2$, leading to $M=25$. This appears to be the optimal choice under the above constraints.
It implies $g_sM^2\gtrsim 500$, which requires an enormous tadpole in the LVS as we shall see in Sect.~\ref{sec:modelimplications}.

Note that from this analysis of Fig.~\ref{fig:poverm}, in order for a metastable minimum to exist in the controlled regime where $g_s<1$ one requires at least $p \geq 2$ anti-D3-branes. Of course additional field strength corrections can alter this result, but assuming this does not occur, the following conclusion is tempting: If $p=1$ the metastable minimum, if one exists, will be at small or zero $\psi$ where the radius of the NS5-brane is string-scale and all orders in $\alpha'$ corrections are important. This object might be thought of as not really an NS5-brane but the single anti-D3-brane, at most smeared out over its Compton wavelength. A single anti-D3-brane then would not polarize into an NS5-brane. This is in line with the logic when one considers the perspective of the nonabelian DBI action of a stack of anti-D3-branes, where the nonabelianity (and hence the presence of more than one anti-brane) is crucial for the Myers effect \cite{Myers:1999ps} to occur.


\subsection{Optimistic bounds on uplifts}
\label{sec:upliftbound}

\begin{figure}[h]
\centering
\includegraphics[width=0.7\textwidth]{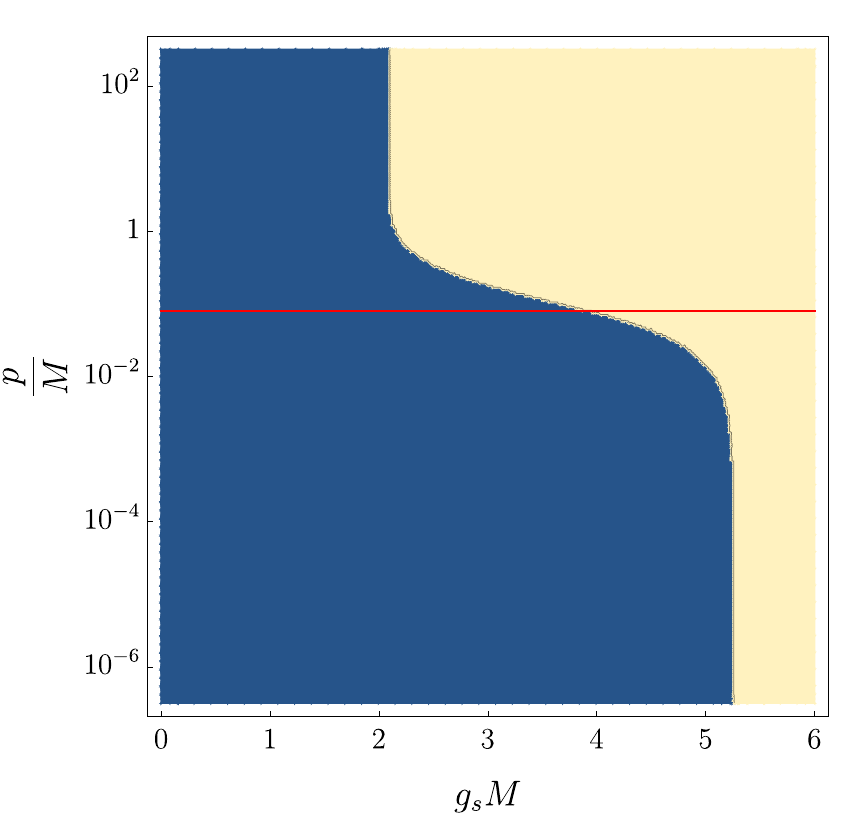}
\caption{The space of curvature-corrected KPV setups, parameterized by $g_s M$ and $p/M$. The regions where the maximum of the potential is at positive/negative value are coloured yellow/blue. For $p/M\ll0.1$ the boundary line asymptotes towards $g_sM\to 5.238$ and for $p/M\gg0.1$ towards $g_sM\to 2.086$. The red line is at $p/M=0.08$, approximately above which the maximum is at stringy or substringy radii for the NS5-brane and hence not trustworthy.}
\label{fig:gsmvpmwheremax}
\end{figure}
We have seen that the $\alpha'^2$-corrected potential \eqref{eq:vtot} always has a maximum, but it has a minimum only in the controlled regime of large $g_sM$.
However, for the purpose of constructing de Sitter vacua, the crucial question is whether our potential $V_\text{tot}$ can provide an uplift. It is not necessary that the corresponding local minimum is in the controlled regime at large $\psi$. Indeed, let us assume that we have a local minimum at small $\psi$, capable of uplifting. The radius of the $S^2$ wrapped by the NS5-brane may then be too small to control the relevant $V_\text{tot}$ against even higher $\alpha'$ corrections. A necessary but not sufficient condition for such an uplift to exist is the presence of a local maximum with a positive potential energy at larger $\psi$. 
We will therefore consider the following very blunt criterion:
We demand that $V_\text{tot}$ has to be positive at its maximum in order to have any chance of providing an uplift. Clearly when the potential is always negative it cannot uplift. As the maximum in the tree level potential $V_{\text{KPV}}$ is always at a fairly central value of $\psi$ it seems reasonable that for part of the parameter space this will also be true for $V_\text{tot}$ and we can neglect even higher order corrections in $\alpha'$ so long as the radius of the tip of the warped throat is not very small in string units. Even then, our criterion can be questioned as we have not accounted for $\alpha'^2$ flux corrections. Still, while the concrete constraints may change in a more complete analysis, our limited analysis should provide an estimate of the scale at which $\alpha'^2$ corrections affect KPV.

The region in $p/M$ and $g_s M$ parameter space where the maximum is positive is shown in Fig.~\ref{fig:gsmvpmwheremax}. One sees that there are two distinct asymptotic regimes: When $p/M\ll0.1$ the bound asymptotes towards $g_sM > 5.238$ and for $p/M\gg0.1$ towards $g_sM > 2.086$. When $p/M \sim 0.1$ there is an interpolating intermediate regime. This clearly suggests that the nature of the maximum of the potential is different in the two asymptotic regimes.

Let us first consider the regime $p/M \ll 0.1$. It is straightforward to evaluate $V_{\text{tot}}(\psi)$ at $p/M=0$. One can convince oneself that the potential has a maximum at $\psi={\cal O}(1)$, where the NS5-brane radius is comparable to the radius of the A-cycle $S^3$. The potential drops sharply away from this maximum. Clearly, the maximum is trustworthy in this regime if the A-cycle radius is sufficiently large.

Now let us consider $p/M \gg 0.1$. We can expand $V_{\text{tot}}(\psi)$ at large $p/M$ and find
\begin{equation}
    V_{\text{tot}}(\psi) = \frac{4 \pi^2}{g_s^3 M} \frac{p}{M}  \left( 2 (g_s M)^2-c_1-  2 c_2 \cot ^2(\psi ) - c_2 \cot ^4(\psi )\right) +\mathcal{O}\left(\left(\frac{p}{M}\right)^0 \right)\,.
\end{equation}
We see that the leading $p/M$ term in this expansion provides a table-shape potential which is essentially flat until $\psi$ gets close enough to the north or south pole such that the $\text{cot}^2 (\psi )$ term begins to compete with the constant $\sim (g_s M)^2$ term. Near $\psi=0$, this happens when $\psi \sim 1 / g_s M$. As a result, the NS5-brane radius is $R_\text{NS5} \sim \sqrt{g_s M} \, \text{sin}(\psi )\sim 1 /\sqrt{g_s M}$. We see that in this region the NS5-brane radius is stringy or sub-stringy in size and hence out of control. The $(p/M)^0$ correction adds a downward slope from $\psi=0$ to $\psi=\pi$. As a result, at large $p/M$ the potential is essentially flat in the entire controlled regime, with a slight downward slope. There is a maximum at the left end of the plateau, near $\psi=0$, which corresponds to stringy or substringy radii for the NS5-brane and can hence not be trusted.{\footnote{ 
We recall that according to \eqref{eq:probeapprox} 
throughout this whole paragraph, even at $\psi\sim {\mathcal O}(1)$, the ratio $p/M$ is restricted by $p/M \ll g_sM$, 
}}

\begin{figure}[h]
    \centering
    \includegraphics[width=\textwidth]{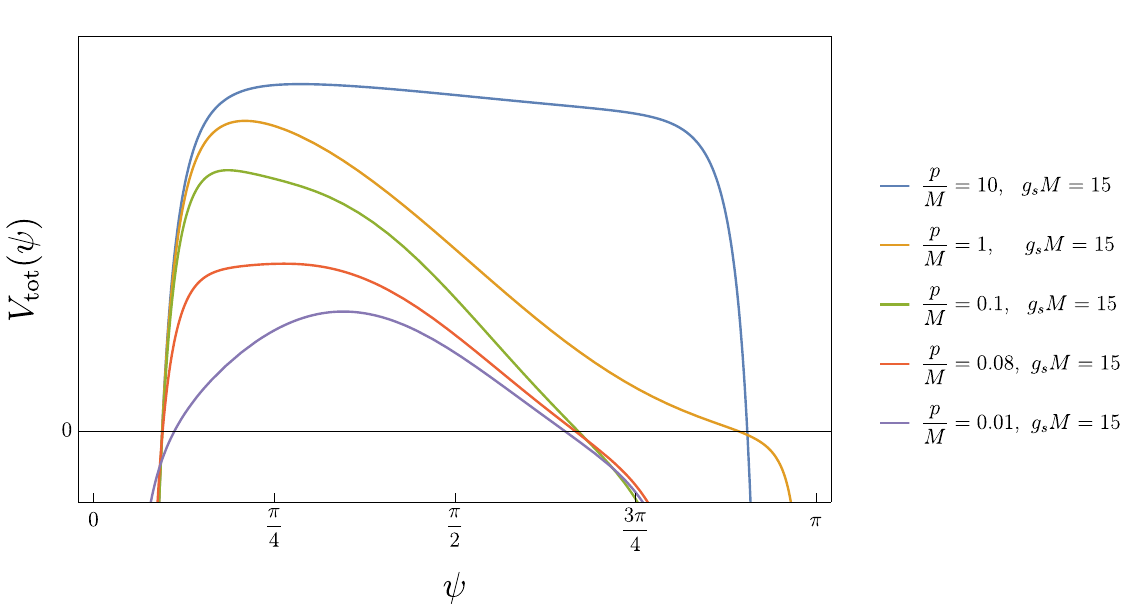}
    \caption{A plot of $V_{\text{tot}}(\psi)$ for different $p/M$ showing the transition between the large and small $p/M$ regime. The potentials at different $p/M$ have been rescaled by choosing $g_s$ to make the comparison of their shape clearer.}
    \label{fig:maximumtransition}
\end{figure}

Somewhere in the intermediate regime $p/M \sim 0.1$ we then lose confidence in our maximum. This transition in where the maximum occurs can already be seen in Fig.~\ref{fig:KPVpsiplot}: As $g_sM$ falls, the maximum coming from the curvature corrections grows, erasing the KPV local minimum. Moreover, we see that depending on whether we make $p/M$ small or large, this maximum emerges at either larger or smaller $\psi$. To make this even more apparent, we have plotted the potential for a wider range of $p/M$ in Fig.~\ref{fig:maximumtransition}. Here we chose $g_sM=15$, a value which is too small for a local minimum to exist.\footnote{{Note that according to \eqref{eq:probeapprox} the probe approximation is only borderline valid for the blue curve. But as we are only trying to visualize the transition from small to large $p/M$, we disregard this issue.}} We still clearly see that, at $p/M \approx 0.08$, a transition in the location of the maximum occurs. Once the maximum has shifted to the left, we lose trust in it since it is now very close to the region where the curvature corrections diverge.

\begin{figure}[h!]
    \centering
    \includegraphics[scale=0.7]{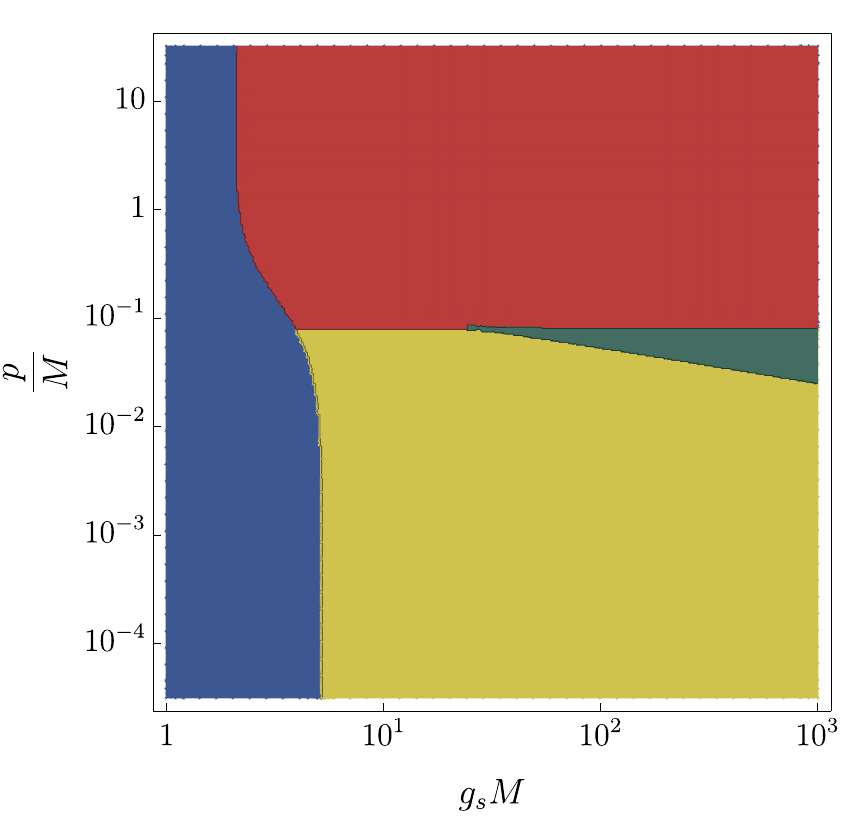}
    \caption{An exclusion plot of the existence of minima and maxima in $V_{\text{tot}}(\psi)$. In the blue region the potential is negative everywhere. In the red region, a positive maximum exists, but it is in an uncontrolled regime. In the yellow region a positive maximum exists in a controlled regime. In the green region a metastable uplifting minimum exists.}
    \label{fig:pMvgMeverythingexclusion}
\end{figure}

We can then also display the regions where a controlled/uncontrolled maximum exists as well as the region with a local minimum of $V_{\text{tot}}$ in a single plot, cf.~Fig.~\ref{fig:pMvgMeverythingexclusion}.
We recall that a controlled maximum implies the presence of a local, SUSY-breaking minimum at smaller $\psi$.

\subsection{Uplifting without deep throats?}
\label{shallowuplift}
When $\psi$ is sufficiently small, $\psi \sim 1/\sqrt{g_s M}$, the NS5-brane radius $R_\text{NS5}\sim \sqrt{g_sM}\sin\psi$ becomes of stringy size. Then even higher order $\alpha'$ corrections become important and we can no longer trust our potential $V_{\text{tot}}(\psi)$. Despite this, we can still make some reasonable estimates of the fully corrected potential at small $\psi$. It is reasonable to assume that at $\psi=0$ the potential matches the potential for $p$ anti-D3-branes. Up to order $\alpha'^2$ in curvature corrections this is given by
\begin{equation}
   p V_{\overline{D3}} = p \mu_3 \text{e}^{-\phi}\left[ 1 - \frac{(4 \pi^2 \alpha')^2}{384 \pi^2} R_{a \alpha b}^{\quad \alpha} R_{\;\beta}^{a \;\; b \beta}\right] = p \mu_3 \text{e}^{-\phi}\left[ 1 - \frac{c}{(g_s M)^2}\right]\,,
   \label{eq:mainantibranecurvecorr}
\end{equation}
with $c=5.9241$. If the fully corrected potential has a maximum but no metastable minimum at intermediate $\psi$ where $V_{\text{tot}}(\psi)$ can be trusted and is set to a finite value at $\psi=0$ given by \eqref{eq:mainantibranecurvecorr} then the fully corrected potential must have a metastable minimum at $\psi$ small (or $0$). It then seems a reasonable estimate that the potential at this minimum is given roughly by \eqref{eq:mainantibranecurvecorr} plus small corrections\footnote{This assumes that the potential rolls down a small amount between $\psi=0$ and the metastable minimum at small $\psi$ as happens in the original KPV set-up shown in Fig.~\ref{fig:KPVpot}. While this seems reasonable, this need not be the case as we are ignorant of the effects of additional corrections. It is e.g. possible that at $p=1$ there is a barrier between $\psi=0$ and a metastable minimum at small $\psi$ and this from the NS5-brane perspective prevents the anti-D3-brane from puffing up classically into an NS5-brane even though both states have the same quantum numbers and transitions between them should be possible, though now heavily suppressed due to quantum tunnelling, due to the no global symmetries conjecture as discussed in \cite{Gao:2020xqh}.}.

It seems that as $\alpha'^2$ curvature corrections contribute negatively to the potential, they can turn the potential negative. This allows for a novel but speculative alternative uplifting mechanism. Suppose that for some parameters the fully corrected potential has a metastable minimum with a positive vacuum energy, while for different parameters (with increased curvature corrections) there exists a metastable minimum with a negative vacuum energy. Then by tuning $g_s$ it should be possible to obtain a metastable minimum with a positive but hierarchically small vacuum energy by using the tension-lowering effect of curvature corrections, without needing a large throat with hierarchical warping.

To obtain a hierarchically small uplift, one requires a finetuning of $g_s$. However, due to the flux dependence of $g_s$ and the density of the string theory flux landscape we believe this finetuning to be no worse than e.g.~the finetuning in $W_0$ required to get a cosmological constant of real world scale.

An issue in estimating exactly when this uplifting mechanism occurs is that when in \eqref{eq:mainantibranecurvecorr} the $\alpha'^2$ corrections are important enough to compete with the leading term, one expects even higher order $\alpha'$ corrections to also be important.

One may take this logic one step further and forget about our warped throat set-up: consider an AdS compactification with small cycles, e.g.~the models of \cite{Demirtas:2021nlu,Demirtas:2021ote}, and place an anti-D3-brane. The anti-D3-brane will be drawn to the small cycle as this lowers their tension due to curvature corrections. If many models with many such small cycles exist, it may by chance happen that for some of them curvature corrections make the anti-brane tension positive but very small such that it is suitable for an uplift. Of course, all this is very speculative and it takes one to (and possibly beyond) the edge of a controlled regime.

{
Let us try to address some of the most immediate concerns which one might have with the control of compactifications relying on our new uplifting mechanism:

Using this novel uplifting mechanism in the LVS, one may worry that the necessary tuning of $g_s$ will affect the LVS AdS minimum in such a way that $|V_\text{AdS}|\approx |V_\text{uplift}|$ can not be satisfied at large volume. However, we see from \eqref{eq:antibranecurvecorr} that near $g_s\sim \sqrt{c}/M$ the potential $V_\text{uplift}(g_s)$ is a growing function, changing sign at $\sqrt{c}/M$. By contrast, $V_\text{AdS}\sim- g_s^{1/2}\text{e}^{-3/g_s}$ asymptotes to zero at an exponential rate. This implies that the sum of these functions will always go through zero for some value of $g_s^\ast$ just slightly above $\sqrt{c}/M$ . Thus, the value of $g_s^\ast$ can be lowered by choosing a larger $M$, such that a large CY volume appears to be easy to achieve.

Furthermore, one might worry that string-scale curvature at the tip implies large curvature also at the mouth of the throat, affecting the rest of the CY and leading to corrections to the Kahler potential. Parametrically, the radius at the tip is given by $R_{S^3}\sim \sqrt{g_s M}$ and the radius at the mouth by $R_\text{mouth}\sim (g_sN+(g_sM)^2)^{1/4}$. Recalling that $N/g_sM^2$ is the exponent governing the warp factor, we have $R^4_{S^3}/R^4_\text{mouth}\approx g_sM^2/N \ll 1$ in the regime of strong warping. At moderate warping, the suppression of the curvature at the mouth is of course limited. Nevertheless, it is parametrically consistent to assume that only the curvature at the tip is string scale.
This goes together the general expectation that localized effects in a CY decouple from the bulk.\footnote{{This sequestering is studied for the KS throat in \cite{Kachru:2007xp,Berg:2010ha}.}}
}

{ There is also the potential danger of loop corrections coming from fields propagating in the throat. Since the geometry near the tip is string scale, a 10d EFT analysis is not reliable. The leading closed-string loop diagram contributing to the potential of the anti-D3-brane is a torus with two vertex insertions, one coupling to the anti-D3-brane, the other corresponding to the graviton background. Correspondingly, at the open-string level there is the annulus with a graviton insertion. Compared to the tree level $\alpha'$ corrections, these diagrams are suppressed by $g_s^2$ and $g_s$. Hence, in the regime $g_s\ll1$ and $g_sM\sim \mathcal{O}(1)$, loop corrections are smaller than the curvature corrections computed above.}

{ Another potential concern is that, as the A-cycle shrinks, branes wrapped on it become light. In our context, such branes are relevant if they additionally fill out from zero to four of the non-compact dimensions. In type IIB, only D3- and D5/NS5-branes meet these criteria. They correspond to particles (cf.~the discussion in \cite{Montero:2021otb}) and domain walls in 4d. Given a string-scale A-cycle, their mass or tension is small -- set by the warped-down string scale. But they are in general not mass- or tensionless, so the EFT does not break down. Thus, while they may play an interesting phenomenological role, this does not directly affect the minimum of the potential and does not threaten the moduli stabilization scenario. A key exception are NS5s since the corresponding domain walls mediate the decay of the SUSY-breaking anti-D3 branes. This is, of course, the main subject of the present paper.}

{
Finally, while the effects discussed in the last three paragraphs can all contribute terms to the Kahler potential which mix Kahler and complex-structure moduli, we do not expect this to be deadly for the proposed scenario. First, the warped-down string scale is higher than the Kahler-moduli mass scale, so complex structure moduli may be integrated out first. Second, since the KS throat does not contain local 2- or 4-cycles, its geometry measured in string units is insensitive to Kahler deformations of the CY. Thus, we expect $\alpha'$ corrections to the Kahler moduli Kahler metric not to be enhanced by the high curvature present in the throat region.
}


\subsection{Open issues}
\label{sec:outlook}

In this paper we have given only an estimate of how $\alpha'^2$-corrections affect the KPV uplift. Let us briefly outline what we believe are the main points which must be addressed in a research program to fully understand $\alpha'$-corrections to KPV.
\begin{itemize}
    \item We have assumed that the $\alpha'^2$ curvature corrections for the NS5-brane in KPV are given by the natural generalization from the known curvature terms on the D5-brane.
    If instead we consider the S-dualized KS set-up \cite{Gautason:2016cyp}, we can be certain that we have included all pure curvature corrections in our analysis.
    It is important to explicitly derive the $\alpha'^2$ curvature terms for the NS5-brane to be sure that they exist and to compute their exact form. Some discussion of curvature terms on NS5-branes not in the DBI-action but related to anomaly cancellation has appeared in \cite{Witten:1996hc,Freed:1998tg,Lechner:2001sj,Grimm:2015mua}. For D-branes, at order $\alpha'^2$, it is known that besides the pure curvature corrections there also exist pure field strength and field strength-curvature mixing terms in the DBI action and some of these have been computed in \cite{Garousi:2009dj,Garousi:2011fc,Garousi:2017fbe,Garousi:2022rcv,BabaeiVelni:2019ptj}. One would expect that by S-duality analogous terms exist for the NS5-brane. It would be crucial to derive these and compute their impact on the KPV potential.
    \item Given the danger posed by the $\alpha'^2$-corrections, one should convince oneself that at $\psi$ where we claim to trust the $\alpha'^2$ corrected potential $V_{\text{tot}}(\psi)$, even higher order $\alpha'$ corrections are sufficiently small and can be neglected. Particularly worrying is that, when the number of anti-branes is one or a fixed small number, the leading order analysis of KPV seems to indicate that the metastable minimum is always such the the NS5-brane wraps a string-scale cycle, making it crucial to know the $\alpha'$-corrections to all orders to know the value of the potential at the metastable minimum (and particularly whether this value is positive). Note that there is some indication that backreaction effects can push the minimum to larger $\psi$ \cite{Cohen-Maldonado:2015ssa}.
    \item Understanding the backreaction of the NS5-brane has played a crucial role in discussions of KPV \cite{Bena:2009xk,McGuirk:2009xx,Bena:2010gs,Bena:2011wh,Bena:2012ek,Bena:2012tx,Bena:2012vz,Bena:2013hr,Blaback:2011nz,Blaback:2011pn,Blaback:2014tfa,Gautason:2013zw,Giecold:2013pza,Blaback:2012nf,Danielsson:2014yga,Michel:2014lva,Cohen-Maldonado:2015lyb,Polchinski:2015bea,Cohen-Maldonado:2015ssa,Cohen-Maldonado:2016cjh,Blaback:2019ucp,Nguyen:2021srl,Bena:2014bxa,Bena:2014jaa,Gautason:2015ola,Bena:2015kia,Armas:2018rsy}. Our discussion of $\alpha'^2$ effects has remained at a probe level. It is crucial to understand the interplay between backreaction and higher-derivative effects.
    \item It would be valuable to understand $\alpha'^2$-effects from the perspective of a stack of nonabelian anti-D3-branes. This calculation should provide an $\alpha'^2$ corrected value of the position of the metastable minimum. Of particular interest is the following issue: As we saw, terms related to the extrinsic curvature play a crucial role in the NS5-brane potential. As the anti-D3-branes are pointlike in the internal dimensions, they have no extrinsic curvature. How then do NS5 extrinsic curvature effects manifest from the perspective of a stack of nonabelian anti-D3-branes? Further, even for the intrinsic curvature, the tensor structure of the terms for the NS5-brane does not match that for the anti-D3-branes. Taking into account all $\alpha'^2$-corrections on the NS5 including pure flux and mixed curvature-flux corrections, we propose the following resolution: For nonabelian brane actions, there is no clear distinction between higher terms in commutators and higher derivative terms \cite{Myers:1999ps,Wyllard:2000qe,Wyllard:2001ye}. Those higher derivative curvature terms of the abelian NS5-brane which do not match on to higher derivative terms of the anti-D3-brane stack we then propose match higher commutator terms in the nonabelian theory. It should be possible to check this by performing the nonabelian analysis of KPV \cite{Kachru:2002gs} at higher order in the commutators. As an example where higher commutator terms can be reinterpreted as higher curvature terms, it is known that a stack of flat nonabelian T-branes can be reinterpreted as a curved abelian brane with an $R^2$ curvature term in the worldvolume action \cite{Bena:2019rth}. As usual, it is unclear how or even if a single anti-D3-brane puffs up into an NS5-brane.
    
    \item It would be interesting to understand impact of the $1/(g_s M)^2$-curvature corrections on the vacua at $\psi=0$ and $\psi=\pi$ from the perspective of the holographic dual field theory, especially as $g_s M$ is the 't Hooft coupling in the dual Klebanov-Strassler field theory \cite{Klebanov:2000hb} and the highly curved regime is precisely the weakly coupled holographic field theory regime. An issue here is that with anti-D3-branes present the existence and precise location of the nonsupersymmetric metastable field theory vacuum corresponding to the presence of anti-branes, as proposed in KPV \cite{Kachru:2002gs}, remains to our knowledge poorly understood\footnote{See e.g.~\cite{DeWolfe:2008zy,Bertolini:2015hua,Krishnan:2018udc} for work on the field theory perspective of such a metastable vacuum under the assumption that it exists.}. In fact, if one would be able to establish the metastable anti-brane vacuum holographically one would be confident that the metastable vacuum indeed exists. The lack of such a holographic field theory understanding of the metastable anti-brane vacuum is thus one of the factors leading to the backreaction discussion \cite{Bena:2009xk}. However, our curvature corrections also affect the situation where no anti-D3-branes are present, the situation described from a holographic field theory perspective in Sect.~2.3 of KPV \cite{Krishnan:2018udc}. In this case, the dual field theory has a baryonic branch, corresponding to a gravity solution with only flux, and a mesonic branch, corresponding to a gravity solution with $M$ D3-branes at the tip of the warped throat \cite{Seiberg:1994bz,Klebanov:2000hb,Kachru:2002gs}. Both branches are supersymmetric, however their moduli spaces are disconnected. There is then a supersymmetric domain wall interpolating between the baryonic and mesonic vacuum, corresponding to the NS5-brane wrapping an $S^2$ being pulled over the A-cycle $S^3$ from $\psi=0$ to $\psi=\pi$ in the gravity picture. According to our curvature-correction computations, when $p=0$, the NS5-brane potential is everywhere negative as soon as $g_sM < 5.238$ (cf.~Fig.~\ref{fig:gsmvpmwheremax}). This seems surprising given that for the flux-only (baryonic) supersymmetric vacuum there are no branes present to provide a negative curvature contribution to the potential. It is possible that additional corrections render the potential positive, however it could also be that for the flux-only vacuum $\alpha'$ curvature corrections in the bulk supergravity action render the potential negative e.g. as discussed in \cite{Junghans:2022exo} the bulk $R^4$ term of \cite{Becker:2002nn} can be corrected by warping and hence lead to terms involving $g_s M$. It would then be worthwhile to understand this in the holographic gauge theory at $g_sM \ll 1$ where our (uncontrolled) calculation claims that the potential should everywhere be very negative.
\end{itemize}

We intend to deal with with some of these important points in a detailed analysis in future work.

\section{Implications for model-building}
\label{sec:modelimplications}

In this section we want to collect some interesting phenomenological applications of the results of Sect.~\ref{sec:KPVcorrected}.

So far we have been concerned with the effects of $\alpha'^2$-curvature corrections to the KPV NS5-brane potential \textit{per se}. The reason we began to study this topic was to better understand the anti-brane uplift in the LVS. With our newfound understanding one can now ask how $\alpha'^2$ corrections to KPV affect string-phenomenological model building i.e.~models such as LVS and KKLT. 

Since our results suggest it might be possible to uplift without a deep throat, it is interesting to ask this question not just for the LVS, but also for e.g. KKLT \cite{Kachru:2003aw}. Since the difficulty of embedding a deep throat in the bulk Calabi-Yau posed the main difficulty in uplifting KKLT \cite{Carta:2019rhx, Gao:2020xqh} (see also \cite{Freivogel:2008wm})
one can ask if our results allow one to resurrect the idea of anti-brane-uplifted KKLT in regimes that previously seemed out of control.

In addition, particles and domain walls in the four external directions have been constructed by wrapping branes on the cycle at the tip of the warped throat \cite{Montero:2021otb,Danielsson:2022odq}. It would be interesting to see how these analyses are affected by curvature corrections of the type considered here.

As each string-phenomenological model comes with many subtleties and additional effects of its own, unrelated to the story of KPV, we will not discuss these in detail here. We will merely state some suggestive results related to the LVS and leave our results on the string-phenomenological impact of $\alpha'^2$ to an upcoming paper \cite{toappear}.

\begin{figure}[h]
    \centering
    \includegraphics[width=\textwidth]{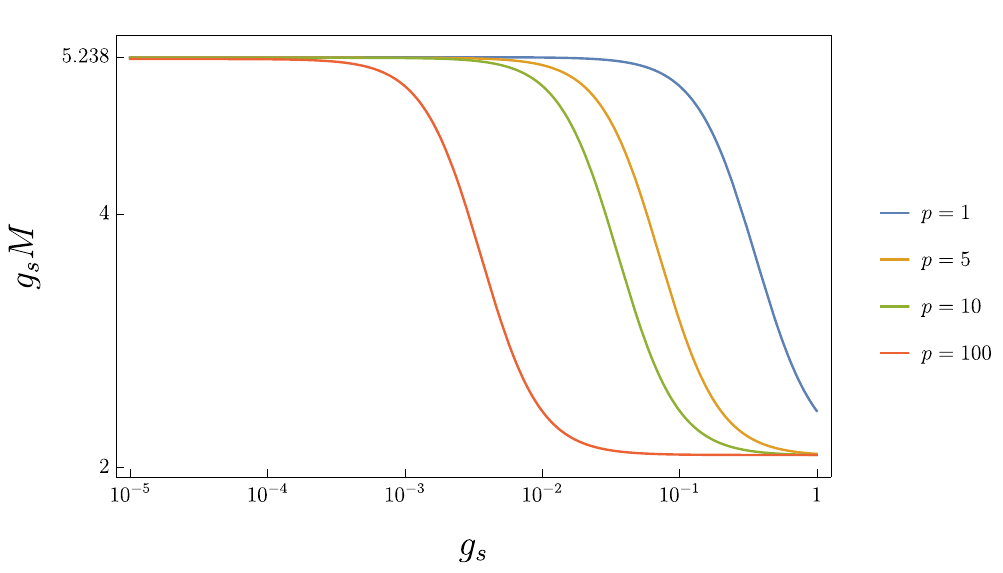}
    \caption{A plot of $g_sM$ over $g_s$ for different values of $p$. The region above the curves is where $V_\text{tot}$ has a positive maximum. For $g_s\ll1$ and $g_s\gg1$ the curves asymptote again towards $g_sM\to 5.238$ and $g_sM\to 2.086$, respectively. Note that when $p/M>0.08$ this maximum is not in a controlled regime.}
    \label{fig:movergsdifferentp}
\end{figure}

The natural parameters in which to understand curvature corrections to KPV are $g_s M$ and $p/M$. If we use KPV in model-building, other parameters will become important. In particular, the individual values of $g_s$, $M$, and $p$ will become important. It is straightforward to display the bounds we have derived in terms of these parameters. In Fig.~\ref{fig:movergsdifferentp} we show the region where $V_{\text{tot}}(\psi)$ has a positive maximum in terms of $g_s M$ as a function of $g_s$ for various $p$.

Given these bounds, we can now compute for the LVS how they constrain the required negative D3 tadpole $-Q_3$ present in the compactification geometry by using the PTC \cite{Gao:2022fdi}. We will present a detailed analysis in \cite{toappear} and merely state the required tadpole for some reasonable parameters here. The additional parameters present in LVS are $a_s$ related to nonperturbative corrections to the superpotential, $\kappa_s$ related to the triple intersection number of the small divisor, $\xi = \chi\, \zeta(3)/ 2 (2 \pi)^3$ where $\chi$ is the Euler number of the Calabi-Yau orientifold, and $c_N$ which is a control parameter related to a bulk curvature correction with varying warp factor. $c_N$ must be large for this correction to be small. Not all the parameters are independent. On way of thinking about this is setting $a_s$, $\kappa_s$, $\chi$, $p$ and $c_N$ will fix $g_s$. This in turn fixes $\tau_s$, the minimal value of $M$ via Sect.~\ref{sec:KPVcorrected}, the volume $\mathcal{V}$ and the minimal value of the D3 tadpole $N$ of the warped throat.

Here, we will proceed differently by choosing a value of $g_s$ (and hence a value of $M$ which is in line with Fig.~\ref{fig:pMvgMeverythingexclusion}) instead of $\xi$. We then use the following equations\footnote{The equations are taken from \cite{Gao:2022fdi}, see equ.~(2.4), (3.2), (3.11) and (3.15) therein.} to determine $\xi$, $\mathcal{V}$ and $N$:
\begin{gather}
    \label{eq:vts}
    \mathcal{V}=\frac{3\kappa_s^{2/3}\xi^{1/3}|W_0|}{2^{7/3}a_s |A_s|g_s^{1/2}}\text{e}^{a_s\frac{\xi^{2/3}}{(2\kappa_s)^{2/3}g_s}}\,,\qquad\qquad
    a_s \frac{\xi^{2/3}}{(2\kappa_s)^{2/3}g_s}=  \frac{16\pi N}{9 g_s M^2}\,,\\
    \label{eq:ptc}
    N  = - \frac{21g_s M^2}{16\pi} \mathcal{W}_{-1}(x)\,, \qquad x =- \frac{3^{11/35}\,a_0^{2/7}}{7\,\,2^{59/105}\, 5^{5/7}\, \pi^{1/35}} \, \frac{\kappa_s^{2/7}}{p^{2/7}\, a_s^{3/7}\, c_N^{5/7}\, (g_sM^2)^{1/7}}
    \,.
\end{gather}
Here, the first equation in \eqref{eq:vts} is the on-shell solution for the volume modulus in the LVS, the second follows from requiring the uplift to be of the same size as the AdS minimum and \eqref{eq:ptc} is called the PTC generalized to generic $p$. 

With this, let us consider some examples where we choose $a_s=2\pi$ as this minimizes the tadpole compared to the choice $a_s=\pi/3$. 

\begin{table}[!h]\centering
	\caption{Minimal value of the required tadpole $(-Q_3)_\text{min}$ for the minimum of $V_\text{tot}$ to occur in the controlled regime for different $p$.
	}
	\vspace{.3cm}
	\label{tab:tadpolecontrolledsec2}
		\begin{tabular}{cccccccccc}
		\toprule
			$p$ & $a_s$ & $\chi$ & $\kappa_s$ & $c_N$ & $g_s M$ & $p/M$ & $g_s$ & $g_s M^2$ & $(-Q_3)_\text{min}$ \\
			\hline
			\midrule
			$2$& $2\pi$&$405$ &$0.1$ &$5$ & $20.58$&$0.0824$ &$0.848$ &$500$ &$1913$\\
			\midrule
			$3$& $2\pi$&$225$ &$0.1$ &$5$ & $20.4$&$0.0825$ &$0.561$ &$742$ &$2900$\\
			\midrule
			$8$& $2\pi$&$56$ &$0.1$ &$5$ &$20.51$ &$0.0824$ &$0.211$ &$1991$ &$8175$\\
			\bottomrule
	\end{tabular}
\end{table}

As can be seen from Table \ref{tab:tadpolecontrolledsec2}, requiring that a controlled metastable minimum exists via the bound in Fig.~\ref{fig:poverm} leads to a minimum value for the tadpole at $\mathcal{O}(10^3-10^4)$. 

\begin{table}[!h]\centering
	\caption{Minimal value of the required tadpole $(-Q_3)_\text{min}$ for $a_s=2\pi$, $p=1$ and different choices of $p/M$, $c_N$ and $g_sM$ demanding the maximum of the corrected KPV potential to be positive.}
	\vspace{.3cm}
	\label{tab:tadpoleinstantonsec2}
		\begin{tabular}{cccccccccccc}
		\toprule
		    \multicolumn{3}{c}{input parameters} & & & & &\multicolumn{2}{c}{$c_N=5$} & & \multicolumn{2}{c}{$c_N=100$}\\
		    \cmidrule(lr){8-9}  \cmidrule(lr){11-12}
			 $g_s M$&$p/M$ &$\kappa_s$&&$g_s$&$g_s M^2$& & $\chi$ & $(-Q_3)_\text{min}$ && $\chi$ & $(-Q_3)_\text{min}$ \\
			\hline
			\midrule
			5& 0.04&  0.1& & 0.2 & 125 & & 43 & 455&&62&580\\
			\midrule
			10& 0.04 & 0.1& & 0.4 & 250& & 124 &921 && 178&1172\\
			\midrule
			15& 0.04& 0.1& & 0.6& 375& & 231& 1395& & 329&1766\\
			\midrule
			20& 0.04& 0.1& & 0.8& 500& & 358&1868& & 510&2365\\
			\midrule
			3.9& 0.08 & 0.1 & & 0.31 & 48.9 & & 82 & 175 & & 118 & 223\\
			\bottomrule
	\end{tabular}
\end{table}

Demanding only that a maximum with a positive potential energy exists leads to much weaker constraints, shown in Table \ref{tab:tadpoleinstantonsec2} where we choose $p=1$. In the last line we show the `most optimistic' setting, i.e.~the point where the blue, red, and yellow region meet in Fig.~\ref{fig:pMvgMeverythingexclusion}. This point is at the boundary of the maximum being positive in a somewhat controlled regime.
Clearly, all numbers should be viewed as an estimate as flux and possible further curvature corrections at order $\alpha'^{2}$ need to be taken into account.

\section{Conclusions and Outlook}
\label{sec:conclusions}

In this paper, we have studied the effects of expected $\alpha'^2$-curvature corrections to KPV. We have seen that these can radically alter the leading-order KPV story, where the potential has a metastable, positive-energy minimum. However, we have only been able to analyze a subset of corrections -- many $\alpha'$-corrections in the NS5-action remain unknown. The main message then is that in this paper we have opened the Pandora's Box of higher-derivative corrections to KPV, but it is not at all clear at the moment where a complete analysis will lead and whether the final outcome at intermediate $g_s M$ will be anything recognizably like the leading-order story of KPV. The different regimes expected on the basis our (partial) $\alpha'^2$-corrections to KPV are summarized in Fig.~\ref{fig:pMvgMeverythingexclusion}, parametrized by $p/M$ and $g_s M$.

We have then used the resulting bounds on $g_s M$ from $\alpha'$-corrections to KPV in the context of the LVS to estimate the minimal negative D3 tadpole. To achieve this, we have employed the Parametric Tadpole Constraint of \cite{Gao:2022fdi}.
To be precise, we have used two different options for how to bound $g_s M$:

The conservative option is to demand that $g_s M$ is sufficiently large such that $\alpha'^2$-corrections to the KPV potential are small, a metastable uplifting minimum exists in the potential, and the leading order story of KPV holds unmodified. How this constrains $g_s M$ can be seen in Fig.~\ref{fig:poverm}. As a result of this constraint, the LVS requires geometries with a negative D3 tadpole of the order of thousands or tens of thousands and at least two anti-D3-branes for the uplift (cf.~Table \ref{tab:tadpolecontrolledsec2}).

The more optimistic approach is to only demand that a local maximum for the NS5-potential exists at positive vacuum energy, such that the standard KPV decay process is classically forbidden. Clearly, this maximum has to be at sufficiently large radius of the $S^2$ wrapped by the NS5, such that our computations are controlled against even higher order $\alpha'$-corrections. 

Given the existence of this maximum one then assumes that, on its SUSY-breaking side, a local minimum capable of uplifting also exists. One accepts that, at the minimum, the NS5-brane wraps a string scale $S^2$, where our computations are not controlled. This constrains $g_s M$ as shown in Fig.~\ref{fig:gsmvpmwheremax} and, in the LVS context, only requires a negative D3 tadpole of the order of hundreds, as summarized in Table \ref{tab:tadpoleinstantonsec2}.

In fact, there is even further cause for optimism: In the analysis described as the optimistic approach above, we have assumed that the potential of the $\alpha'$-corrected minimum is at the same energy scale as the leading-order anti-D3-brane uplift. If, however, the $\alpha'^2$-corrections significantly alter the NS5-brane potential, it is entirely plausible that the metastable minimum could be at much lower energy. It may even change sign, leading to non-SUSY AdS. This is in fact the situation which a pure anti-D3-analysis (without transition to the NS5) suggests \cite{Junghans:2022kxg}. Now, if one can tune the uplifting potential to extremely small values, one may be able to produce an exponentially suppressed uplift without relying on a deep warped throat. This tuning could be realized using the flux discretuum of $g_s$, which enters the $1/(g_s M)^2$ prefactor of curvature corrections. If, as a result, one no longer requires a deep throat, the main difficulty in achieving uplifted KKLT or LVS vacua disappears\footnote{All this is a concrete example of the general philosophy of how one expects the interesting vacua with all moduli stabilized to live in the interior of parameter space rather than at the asymptotics \cite{Dine:1985he}. The more orders in corrections we take into account, the richer the potential will be and the more options there are for model-building. If it is possible to explicitly compute all important $\alpha'^2$-corrections, we should do so and embrace the added richness this gives us.}.

Although we have not discussed this in detail, one could perform an analogous analysis in the S-dual set-up to KPV where the anti-D3-branes puff up into a D5-brane at the tip of the throat. It would be interesting to treat this set-up in full detail.

The issue with making the preceding ideas more concrete is that not all $\alpha'^2$-corrections to KPV are explicitly known. As an outlook for future research, let us summarize again which points we believe need to be understood to obtain a clear, controlled picture of $\alpha'$-corrected KPV:
\begin{itemize}
    \item The $\alpha'^2$-corrected NS5-brane action has to be computed explicitly, both curvature terms and gauge field strength contributions. As we have not included the field strength terms at all, our results may change significantly. This task should be easier in the S-dualized KS set-up as more is known about $\alpha'^2$ corrections on the D5-brane.
    \item It would be important to check that, at least near the maximum of the corrected potential, the $\alpha'^2$ effect is trustworthy in the sense that even higher order $\alpha'$-corrections are small and can be neglected. 
    \item It is crucial to include backreaction effects, { especially the notoriously difficult flux backreaction, in our analysis.}
    \item In regimes where the $S^2$ wrapped by the NS5 is expected to be of stringy size, it would be interesting to consider the anti-D3-brane picture and compute higher-commutator and higher-derivative effects up to a certain order in $\alpha'$ to estimate if and at which angle $\psi$ the minimum exists.
    \item It would be desirable to develop a deeper understanding in the holographically dual field theory of the vacua on both sides of the domain wall which corresponds to moving the NS5-brane from the south to the north pole. A natural starting point would be to study the case without anti-branes, where both the baryonic and mesonic branch are supersymmetric.
\end{itemize}

\newpage
\section*{Acknowledgements}
We thank Thomas Van Riet, Vincent Van Hemelryck, and Daniel Junghans for valuable discussion.

This work was supported by the Graduiertenkolleg ‘Particle physics beyond the Standard Model’ (GRK 1940) and the Deutsche Forschungsgemeinschaft (DFG, German Research Foundation) under Germany’s Excellence Strategy EXC 2181/1 - 390900948 (the Heidelberg STRUCTURES Excellence Cluster).

\appendix

\section{Curvature computations}
\label{sec:curvapp}
The curvature tensors in \eqref{r2} are given by \cite{Bachas:1999um}
\begin{align}
    (R_T)_{\alpha \beta \gamma \delta} &= R_{\alpha \beta \gamma \delta} + g_{a b} (\Omega^a_{\alpha \gamma} \Omega^b_{\beta \delta} - \Omega^a_{\alpha \delta} \Omega^b_{\beta \gamma})\,,\\
    (R_N)_{\alpha\beta}^{\quad a b} &= -R^{a b}_{\,\,\,\,\alpha \beta} + g^{\gamma \delta} (\Omega^a_{\alpha \gamma} \Omega^b_{\beta \delta} - \Omega^b_{\alpha \gamma} \Omega^a_{\beta \delta})\,,\\
    \overline{R}_{a b} &= \hat{R}_{a b} + g^{\alpha \alpha'} g^{\beta \beta'} \Omega_{a \, \alpha \beta} \Omega_{b \, \alpha' \beta'}\,.
\end{align}
Here, $R^\mu_{\;\nu \rho \sigma}$ is the 10D Riemann tensor, $\mu,..$ indices run over all ten dimensions, $\alpha,..$ indices run tangent to the brane, and $a,..$ indices run normal to the brane. One obtains $\hat{R}_{a b} = R^\alpha_{\; a \alpha b}$ and $(R_T)_{\alpha \beta} = (R_T)^\gamma_{\;\alpha \gamma \beta}$ by contracting only over indices tangent to the brane. The second fundamental form $\Omega^\mu_{\alpha \beta}$ of the brane worldvolume is given by
\begin{equation}
    \Omega^\mu_{\alpha \beta} = \partial_\alpha \partial_\beta Y^\mu - (\Gamma_T)^\gamma_{\alpha \beta} \partial_\gamma Y^\mu + \Gamma^\mu_{\nu \rho} \partial_\alpha Y^\nu \partial_\beta Y^\rho\,,
\end{equation}
with $\Gamma^\mu_{\nu \rho}$ the 10D target space connection, $(\Gamma_T)^\gamma_{\alpha \beta}$ the connection on the induced metric on the brane worldvolume and $Y^\mu (\xi^\alpha)$ the coordinate functions describing the brane embedding in the 10D geometry. We defined $\Omega^a_{\alpha\beta}\coloneqq\left. \Omega^\mu_{\alpha\beta}\right|_{\mu=a}$ such that the normal indices are contracted with $g_{ab}$.

When a brane is embedded in flat space, the nonzero curvature terms are entirely given by terms related to the second fundamental form. When the brane is totally geodesically embedded, the curvature terms are entirely given by terms related to the 10D Riemann tensor.

As the anti-D3-brane is pointlike in the internal geometry, it is totally geodesically embedded. The NS5-brane however is nongeodesically embedded unless it is wrapping the equator of the $S^3$ at $\psi= \pi/2$.

As the metric of the Klebanov-Strassler throat, \eqref{eq:KSmetric}, is known exactly, it is in principle straightforward to compute all the necessary curvature terms. However, in the natural metric for the deformed conifold, \eqref{eq:conifold}, the coordinates of the $S^2$ wrapped by the NS5-brane are not cleanly isolated. To compute the curvature terms we proceed as follows: To obtain the coordinates wrapped by the NS5 we use the spherical parametrization of the deformed conifold at the tip of the throat \cite{Nguyen:2019syc} where the $S^3$ and the $S^2$ are parametrized with standard spherical coordinates $(\psi,\omega,\varphi)$ and $(\theta,\phi)$, respectively. The NS5 hence wraps the $(\omega,\varphi)$ plane inside the $S^3$. The metric is given by 
    \begin{equation}
        \dd s_6^2 = \mathcal{F}(\tau)\, \text{Tr} \left( \dd W^\dagger \dd W \right) + \mathcal{G}(\tau)\, \left|\text{Tr}\left( W^\dagger \dd W \right)\right|^2\,,
    \end{equation}
where, setting $\varepsilon=1$ as above,
    \begin{gather}
        \mathcal{F}(\tau) =\frac{\left( \sinh(2\tau) - 2\tau \right)^{1/3}}{2^{4/3}\sinh\tau}\,,\\
        \mathcal{G}(\tau) = \frac{2-3 \coth^2 \tau+ 3\tau\left( \cosh\tau/\sinh^3\tau \right)}{12 \left( \cosh\tau\sinh\tau-\tau \right)^{2/3}}\,,\\
        W = \left( 1+\frac{\tau^2}{8} + \frac{\tau^4}{384} +\mathcal{O}(\tau^6) \right) L + \left( \frac{\tau}{2}+ \frac{\tau^3}{48} +\mathcal{O}(\tau^5) \right) L \hat{L}\,.
    \end{gather}
Here,
    \begin{gather}
        L =
        \begin{pmatrix}
            -\sin\psi\,\sin\omega\,\cos\varphi+ \text{i} \sin\psi\,\sin\omega\,\sin\varphi & \cos\psi-\text{i} \sin\psi\,\cos\omega \\
            \cos\psi+\text{i} \sin\psi\,\cos\omega & \sin\psi\,\sin\omega\,\cos\varphi+ \text{i} \sin\psi\,\sin\omega\,\sin\varphi\\
        \end{pmatrix}\,,\\
        \hat{L} = 
        \begin{pmatrix}
            -\cos\theta & -\text{e}^{\text{i}\phi} \sin\theta\\
            -\text{e}^{-\text{i}\phi} \sin\theta & \cos\theta\\
        \end{pmatrix}\,.
    \end{gather}
With this, \eqref{eq:KSmetric} and the warp factor \eqref{eq:warpfactor} it is a straightforward but tedious calculation to obtain the $R^2$ corrections at the tip.
We obtain the following results: 
\begin{align}
\label{eq:rns51}
    (R_T)_{\alpha \beta \gamma \delta}(R_T)^{\alpha \beta \gamma \delta}  =& \frac{6^{2/3}}{g_s^2 M^2 I(0)} + \frac{6^{2/3}}{g_s^2M^2 I(0)} \left( \cot^4\psi + 2\cot^2\psi \right)\\
    \label{eq:rns52}
    -2 (R_T)_{\alpha\beta}(R_T)^{\alpha\beta}  = &-2 \, \frac{6^{2/3}}{2 g_s^2 M^2 I(0)} - 2 \frac{6^{2/3}}{2g_s^2M^2 I(0)} \left( \cot^4\psi + 2\cot^2\psi \right) \\
    \label{eq:rns53}
    - (R_N)_{\gamma\delta a b}(R_N)^{\gamma\delta a b}  = & - \frac{6^{2/3}}{g_s^2 M^2 I(0)}
\end{align}
\begin{equation}
        \label{eq:rns54}
    2 \overline{R}_{ab}\overline{R}^{ab}  =  2 \,\frac{3^{2/3} \left( 117\,  I(0)^2- 200 \,I(0) \,I''(0) + 150\, I''(0)^2 \right)}{25 \times 2^{1/3} g_s^2 M^2 I(0)^3} 
     + \frac{2\times 6^{2/3} \cot^2\psi}{g_s^2M^2 I(0)}\left(2+\cot^2\psi\right)
\end{equation}
The terms \eqref{eq:rns51} and \eqref{eq:rns52} cancel against each other. In \eqref{eq:rns54} the term $\sim \cot^2\psi$ stems from the $\overline{R}\Omega^2$ term and the term $\sim\cot^4\psi$ from $\Omega^4$. This also matches the naive expectation of the extrinsic curvature of an $S^2$ which scales as $1/R_{S^2}$. Moreover, for $\psi=\pi/2$ the extrinsic curvature terms vanish as at the equator the NS5 is geodesically embedded. Inserting \eqref{eq:rns51}-\eqref{eq:rns54} into \eqref{rNS5} and integrating over the $S^2$ gives the $\alpha'^{2}$ curvature corrected KPV potential \eqref{eq:vcurv}.

\bibliographystyle{JHEP}
\bibliography{refs}

\end{document}